\documentclass{optica-article}

\journal{opticajournal} 

\articletype{Research Article}
\newcommand{\SI}{supplementary material}

\begin{document}

\title{Nitrogen-Vacancy Color Centers in Nanodiamonds as Reference Single-Photon Emitters}

\author{Nikesh Patel,\authormark{1} Benyam Dejen,\authormark{2} Stephen Church, \authormark{1,*} Philip Dolan\authormark{2,3} and Patrick Parkinson\authormark{1} }

\address{\authormark{1}Department of Physics \& Astronomy and the Photon Science Institute, University of Manchester, Oxford Road, Manchester, M13 9PL, United Kingdom\\
\authormark{2}National Physical Laboratory, Hampton Road, Teddington, Middlesex, TW11 0LW, United Kingdom\\
\authormark{3}Present address: Nu Quantum, Broers Building, 21 JJ Thomson Avenue, Cambridge, CB3 0FA, United Kingdom}

\email{\authormark{*}stephen.church@manchester.ac.uk} 


\begin{abstract*} 
Quantitative and reproducible optical characterization of single quantum emitters is crucial for quantum photonic materials research, yet accounting for measurement conditions remains challenging due to a lack of an established reference standard. We propose nanodiamonds containing single nitrogen vacancy (NV$^{-}$) color centers as reliable, stable and robust reference sources of single-photon emission. We select 4 potential reference emitter candidates from a study of thousands of NV$^{-}$ centers. Candidates were remeasured at a second laboratory, correlating optical pump power and NV$^{-}$ center emission intensity at saturation in addition to corresponding $g^{(2)}(0)$ values. A reference nanodiamond is demonstrated to account for experimental conditions, with reproducible and reliable single-photon emission, as a model for a new, to our knowledge, single-photon emitter reference standard.

\end{abstract*}

\section{\label{sec:level1}Introduction}
Single-photon emitters (SPEs) will play a ubiquitous role in quantum technologies 2.0, motivated by growing interest in applications such as quantum sensing,\cite{Terada2019MonodisperseCenters} computing\cite{Li2023BrightInjections,M.Valensise2022Large-scaleProcessing} and communication.\cite{Liu2013Single-photonEpitaxy} However, realization of triggerable (also referred to as "on-demand" or "deterministic") SPEs for quantum photonic technologies, such as quantum key distribution,\cite{Basset2021QuantumDot} qubit transmission nodes\cite{Bathen2021ManipulatingCarbide} or quantum logic gates\cite{Zhai2022QuantumDots} remains challenging. Potential SPEs include defects in bulk crystals,\cite{Day2023LaserCavities,Li2024HeterogeneousPlatform,Wood2022LongNanodiamonds}, self-assembled quantum dots,\cite{Liu2013Single-photonEpitaxy} colloidal quantum dots\cite{Proppe2023HighlyDots,Feng2017PurificationDots} or quantum dot-in-nanowire systems.\cite{Reimer2012BrightNanowires,Borgstrom2005,Heiss2013Self-assembledPhotonics,Lazi_2017,Loitsch2015CrystalNanowires} CQDs have emerged as a promising SPEs due to their exceptional quantum yield, \cite{Nelson2024} along with perovskite quantum dots with high single-photon purity,\cite{Wang2024} local strain-activated SPEs in transition metal dichalcogenides (TMDs) \cite{Wang2024a,Chen2024} and hexagonal boron nitride (hBN). \cite{Chen2024b} These materials can be further modified by adding optical reflectors,\cite{Haws2022BroadbandReflector} surface passivation or hosting quantum dots in photonic cavities\cite{Adekanye2018RobustMicrocavity,Giovannetti1992Quantum-enhancedLimit} and metasurfaces\cite{Iyer2024ControlMetasurfaces} to enhance the spectral density of emission. However, accurate and reproducible comparison of novel single-photon emitting materials, which often incorporate anisotropic photonic enhancements remains problematic\cite{Esmann2024} due to the lack of a reference with a known single-photon emission rate.
A suitable reference must allow quantitative comparison between multi-photon emission probability as measured by the second-order degree of coherence,\cite{Glauber1963TheCoherence} $g^{(2)}(t)$, single-photon emission rate, and saturation pumping rate. The emission rates of SPEs are limited by their state emission lifetime, and typically exhibit saturation with increased pump intensity. This is observed as a sub-linear relationship between illumination intensity and single-photon emission rate.\cite{Bluvstein2019IdentifyingCenters} Critically, this relationship can be exploited to excite a system at the same rate using different experimental apparatus. This is achieved by setting the pump power to a fixed rate below saturation, which allows reproducible measurements under fixed internal excitation conditions, enabling quantitative comparison of single-photon emitting materials characterized with different experimental approaches.

Nanodiamonds (NDs) are an ideal single-photon emitter (SPE) \cite{Doherty2013TheDiamond,Giovannetti1992Quantum-enhancedLimit} when they contain a single nitrogen-vacancy (NV$^{-}$) color center,\cite{Rodiek2017ExperimentalNanodiamond} known to emit single photons on demand.\cite{Lounis2005Single-photonSources,Eisaman2011,Brouri2000PhotonDiamond,Kurtsiefer2000StablePhotons} They have beneficial SPE characteristics due to stability against photobleaching,\cite{Kurtsiefer2000StablePhotons} ease of fabrication,\cite{Botsoa2011OptimalSpectroscopies,Dantelle2010} and room-temperature emission.\cite{Adekanye2018RobustMicrocavity,Beveratos2002RoomSource,Kurtsiefer2000StablePhotons} NDs are preferred over bulk diamond as the size constraint reduces the probability of hosting multiple NV$^{-}$ defects, and they can have higher single-photon emission rates due to Purcell enhancement.\cite{Purcell1946ResonanceSolid,Zalogina2018PurcellNanoantennas} It has been shown that removal of graphite residues,\cite{Krueger2012FunctionalityNanodiamond} contaminants and surface defects can improve performance.\cite{Hirt2021SampleNanodiamonds} Furthermore, NDs are portable as they can be deposited on many substrates, and the charge-conversion process is unaffected by temperature between 20 and 75~\textdegree{C}. \cite{Barbosa2024} Previous studies have explored their use as reference emitters to calibrate single-photon detectors\cite{Kuck2022SingleState-of-the-art,vonHelversen2019QuantumDetectors} and for a comparison of $g^{(2)}(0)$ measurements.\cite{Moreva2019FeasibilityRange} Here, we extend these approaches by registering single NV$^{-}$ defects and demonstrating their use as a portable reference standard.

The NV$^{-}$ color center is a substitutional nitrogen point defect coupled to an adjacent vacancy site\cite{Giovannetti1992Quantum-enhancedLimit,Doherty2013TheDiamond,Jelezko2006SingleReview} in the diamond lattice. The nitrogen-vacancy center is stable due to the exceptional chemical and mechanical stability of the diamond lattice.\cite{Jung2021} Diffusion of the defect towards the surface occurs on a timescale of centuries.\cite{Laube2023} The emission from fluorescent nanodiamonds (FNDs), used in this work, are photostable,\cite{Chang2008} showing no photobleaching or fluorescent intermittency – even for single NV centers.\cite{Yu2005}

The fluorescent intensities of milled FNDs are bulk-dependent, and their emission is unaffected by surface properties or environment.\cite{Chang2008} However, for detonation nanodiamonds, surface termination species can affect the emission properties, where carboxylated nanodiamonds can reduce the emission by up to 100x compared to other species.\cite{Reineck2017} For this reason, we preferably use milled FNDs.

Two fluorescent charge states exist,\cite{Berthel2015PhotophysicsNanocrystals} NV$^{0}$ and NV$^{-}$, where the latter is optically active with a lifetime between 9-25~ns.\cite{Laube2019ControllingNanodiamonds,Li2015LifetimeNanodiamonds} Under 532~nm optical pumping, a common wavelength for single-photon experimentation in the visible range, the NV center can be ionized into either state (70\% NV$^{-}$ and 30\% NV$^{0}$),\cite{Waldherr2011DarkNMR,Hauf2011ChemicalDiamond} however preferentially favors NV$^{-}$ at longer excitation wavelengths.\cite{Li2015LifetimeNanodiamonds} NV$^{-}$ emits with a zero-phonon line (ZPL) at 637~nm, extending up to 800~nm via phonon sidebands, whereas NV$^{0}$ emits with a ZPL at 575~nm. The preferred state to use as an SPE is NV$^{-}$ as it has a significantly shorter lifetime compared to NV$^{0}$\cite{Han2010} and a higher count rate. Even without spectrally filtering the NV$^{0}$ emission, the majority of emission originates from NV$^{-}$. NV$^{-}$ centers are known to exhibit emission saturation,\cite{Bluvstein2019IdentifyingCenters,Rodiek2017ExperimentalNanodiamond} limited by the lifetime of the excited state.

We validate the use of NV$^{-}$ color centers as single-photon emitter reference standards by measuring the $g^{(2)}(0)$ value, pumping power and count rates at saturation using two systems with different experimental apparatus at room temperature. In this work, 1053 objects were initially examined at the first laboratory (NPL) using a high-throughput quasi-confocal Hanbury Brown and Twiss\cite{BROWN1956CorrelationLight} (HBT) interferometer. This was filtered to 253, according to an intensity threshold which varied depending on the background. This left 111 potential SPE NV$^{-}$ centers with $g^{(2)}(0)$~<~0.5. A subset of 6 NV$^{-}$ candidate centers, labeled as \#A-\#F in the text, from the 111 were selected, due to their favorable signal-to-noise ratio (SNR) and a detailed description of the screening process is given (see \SI). These represent 5.41\% of the total identified SPEs. These 6 candidates were re-examined at a second laboratory, the University of Manchester (UoM), to check whether their properties were recoverable and 4 of the 6 were verified to contain a single emitter. All parameters were measured at both laboratories and remeasured at UoM to assess stability. Measurements were focused on one candidate, which showed repeatability with a 3\% standard deviation in count rate at saturation from the measurements obtained in table \ref{tab:ND5_repeats}. Repeatability is a critical aspect for the NV$^{-}$ centers to serve as a dependable reference, and when coupled with the robust qualities of NV$^{-}$ centers hosted in NDs, we show these SPEs can be used as an efficacious benchmark for single-photon experimentation.

\section{\label{sec:methods}Methods}%

\subsection{\label{sec:methods_spincoating}Spincoating Nanodiamonds}
A silicon substrate was prepared for lithography by cleaning in acetone, isopropyl alcohol and then dried with nitrogen. The silicon was treated in O$_{2}$ plasma for 6 minutes prior to photoresist deposition. A photoresist (AZ 5214E, Merck Performance Materials GmbH) was spin-coated at 4000~rpm for 60~s and baked at 110~°C for 2 minutes, forming a 285~nm thick SiO$_{2}$ surface layer. A pattern was applied by optical grid mark lithography using a Microtech LW-405B+ using a ring focused laser producing an energy density (fluence) of 262~mJ~cm$^{-2}$. The exposed substrate was developed in undiluted AZ~726~MIF for 40~s and sonicated in acetone to remove organic residue. The substrate was ozone cleaned to create a hydrophilic surface. FND Biotech brFND-70/50/35/10 1 mg/ml NDs were diluted in deionized water, sonicated, and then dropped onto the patterned substrate until complete coverage and allowed to settle for 60~s. The sample was spincast at 1000~rpm to eject excess material, leaving an approximately 0.03~${\mu}$m$^{-2}$ isotropic distribution of NDs by analysis of a microscope image (see~\SI).

\subsection{\label{sec:methods_NPLSetup}High-throughput NV$^{-}$ Center Identification}
The initial search for candidate single emitting NV$^{-}$ centers at NPL used a high-throughput confocal microscopy system (see~\SI) with an HBT collection path containing two single-photon avalanche photodiodes (SPADs) and a PicoQuant PicoHarp 300 time-correlated single-photon counting (TCSPC) module.

A 532~nm continuous wave (CW) laser was directed through a 60$\times$ magnification, 0.90 numerical aperture (NA) objective lens by a 550~nm cut-on dichroic mirror, providing a diffraction limited Gaussian spot. The beam was raster-scanned across the sample surface, placed on a platform, in 200~nm steps by using a fast-steering mirror. The NV$^{-}$ emission was additionally filtered by a 550~nm long-pass filter before being focused by a 10$\times$ magnification, 0.26~NA objective into a single-mode patch cable connected to one input fiber of a 50~$\mu$m core multi-mode 2$\times$2 fiber beam splitter whose output fibers were each attached to an Excelitas SPCM-AQR-14-FC SPAD. The signals from the SPADs were sent to a PicoHarp 300 TCSPC module operating in time-tagged time-resolved (TTTR) mode. The count rate maps were expressed in counts per second (cps), with bright spots that correspond to emissive NV$^{-}$ centers. Several 50~$\mu$m~$\times$~50~$\mu$m areas were measured, containing hundreds of NV$^{-}$ centers per scan. A count rate threshold filtered which NV$^{-}$ centers to investigate further, in combination with an object finding algorithm (see~\SI). A background count rate was recorded before proceeding with power dependence measurements on each identified object and its photon stream was acquired for 600~s in TTTR mode at 80\% of the pump power saturation point. Six potential single emitting NV$^{-}$ centers were selected from a pool of 111 potential emitters with $g^{(2)}(0)$~<~0.5 and root-mean square error (RMSE)~<~0.15 in the regions ${\lvert}t{\rvert}>1~\mu$s. The region ${\lvert}t{\rvert}>1~\mu$s was chosen to evaluate the RMSE as it is sufficiently longer than the NV$^-$ lifetime (12~ns) and other longer-lived states (<~500~ns), therefore corresponding only to background noise within the system.

\subsection{\label{sec:methods_UoMSetup}Candidate NV$^{-}$ Center Validation}
The candidate NV$^{-}$ centers were remeasured at UoM (see~\SI) using a 532~nm CW diode-pumped solid state laser which passed through a 532~nm laser line filter. The sample was mounted on a PhysikInstrumente V-738 high-precision XY motor-controlled stage. The laser was focused on the sample by a 100$\times$ magnification, 0.85~NA objective lens with an elliptical beam profile ($d_{min}=560$\,nm, $d_{maj}=740$\,nm). The emission passed through a 550~nm dichroic mirror, a 550~nm longpass filter and two 532~nm Raman notch filters. The filtered emission was focused by an $f=40$\,mm lens on to a $50$\,$\mu$m multimode 1$\times$2 fiber beam splitter, attached to two ID Quantique ID100 SPADs. The SPADs were connected to a PicoQuant HydraHarp 400 TCSPC module. Due to the lower light collection efficiency compared to NPL, TTTR data for each candidate ND was acquired for 3600~s at 80\% of the pump power saturation point. In some cases, this resulted in a SNR that was insufficient for generating g$^{(2)}(t)$ signals - when this occurred, pumping was performed at a higher power, set to 0.1~mW.

\section{\label{sec:Results and Discussion}Results and Discussion}

\begin{figure*}
    \centering
    \includegraphics[width=14cm]{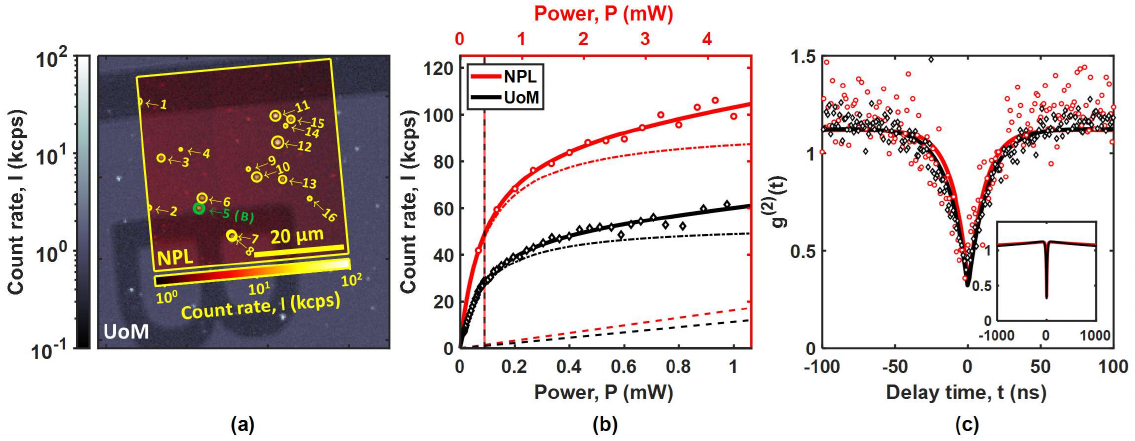}
    \caption{a) Overlayed count rate maps of a region measured for NV$^{-}$ identification from NPL (red) and UoM (gray). The NV$^{-}$ centers examined at NPL are circled. ND\#5 represents ND candidate \#B, circled in green. b) Power dependence plots of \#B measured at UoM (black) and NPL (red). The \textit{vertical} lines correspond to $P_{sat}^{NPL}=0.39(6)$\,mW (solid) and $P_{sat}^{UoM}=0.089(5)$\,~mW (dashed), with the full saturation curve (solid), components from the NV$^{-}$ emission (dashed-dot) and linear residual laser offset (dashed). Note: the x-axes are scaled from 0 to 12$P_{sat}$ such that the vertical lines corresponding to $P_{sat}^{NPL}$ and $P_{sat}^{UoM}$ overlap following SPE system calibration. c) $g^{(2)}(t)$ signals for \#B with the inset covering a larger coherence window to show bunching decay where $g^{(2)}(t)>1$.}
    \label{figure:tcspc_mapping}
\end{figure*}

A total of 1053 objects were initially identified at NPL, a subset of this is shown in FIG. \ref{figure:tcspc_mapping}(a). The 1053 objects were filtered to 253 (24.0\%) by rejecting objects that emit below a minimum count rate threshold (see~\SI). Count rate, $I$, and pump power saturation, $P_{sat}$, were measured as a function of pump power, $P$, modeled using:

\begin{eqnarray}\label{eqn:psat_fit}
    I(P)=\frac{k_{\infty}P}{P+P_{sat}}+cP,
\end{eqnarray}

where $P$ is the incident power measured at the sample surface, $k_{\infty}$ is defined as the maximum observable count rate at infinite pump power,\cite{Kumar2016CouplingNanocube} $P_{sat}$ is the power at half of the maximum observable count rate and $c$ is an experimental parameter accounting for background photon flux as a function of pump power.\cite{Rodiek2017ExperimentalNanodiamond} The NV$^{-}$ count rate at pump power saturation is $I(P_{sat})$=$\frac{k_{\infty}}{2}$. The first term in Eq. (\ref{eqn:psat_fit}) corresponds to emission from the NV$^{-}$ center and the second term is proportional to the laser pump signal. FIG. \ref{figure:tcspc_mapping}(b) shows by setting pump power to 80\% of $P_{sat}$, we ensure that the NV$^-$ center is excited at the same rate at UoM as at NPL. NV$^{-}$ center \#B can be used as a reference single-photon emitter associated with a count rate denoted as $I_{80}$~=~$I(0.8P_{sat})$. This condition allows quantitative comparison between the NV$^{-}$ centers measured at UoM and NPL, therefore $g^{(2)}(t)$ data was acquired at this power. The $g^{(2)}(t)$ signals, in FIG.\ref{figure:tcspc_mapping}(c), were fit by a three-level model\cite{Berthel2015PhotophysicsNanocrystals} using Eq. (\ref{eqn:g2fit}). The first exponential term accounts for the anti-bunching from the NV$^{-}$ centers, with a bunching contribution in the second exponential:

\begin{eqnarray}\label{eqn:g2fit}
    g^{(2)}(t) = 1 - a\left[
    (b){exp}\left(\frac{-|t-t_{0}|}{\tau_{NV}}\right) + (1-b){exp}\left(\frac{-|t-t_{0}|}{\tau_{L}}\right)
    \right]
\end{eqnarray}

where $1-a=g^{(2)}(0)$, $b$ is a weighting factor between the two exponent terms, $\tau_{NV}$ is the anti-bunching lifetime and $\tau_{L}$ represents bunching from longer-lived contributions at finite delay times where $|t|>>\tau_{NV}$. Contributions to $\tau_{L}$ come from shelving to a metastable state\cite{Li2015LifetimeNanodiamonds,Doherty2013TheDiamond,Drabenstedt1999Low-temperatureDiamond,Kurtsiefer2000StablePhotons,Berthel2015PhotophysicsNanocrystals} and charge state switching\cite{Beha2012OptimumDiamond,Aslam2013Photo-inducedDetection,Han2010MetastableResolution} between the NV$^0$ and NV$^-$ states. $t_{0}$ accounts for an arbitrary delay time shift in the $g^{(2)}(0)$ dip, due to a relative time delay between the SPADs, resulting in an offset between photon arrival times at each channel.

\begin{figure}
    \centering
       \includegraphics[width=8.5cm]{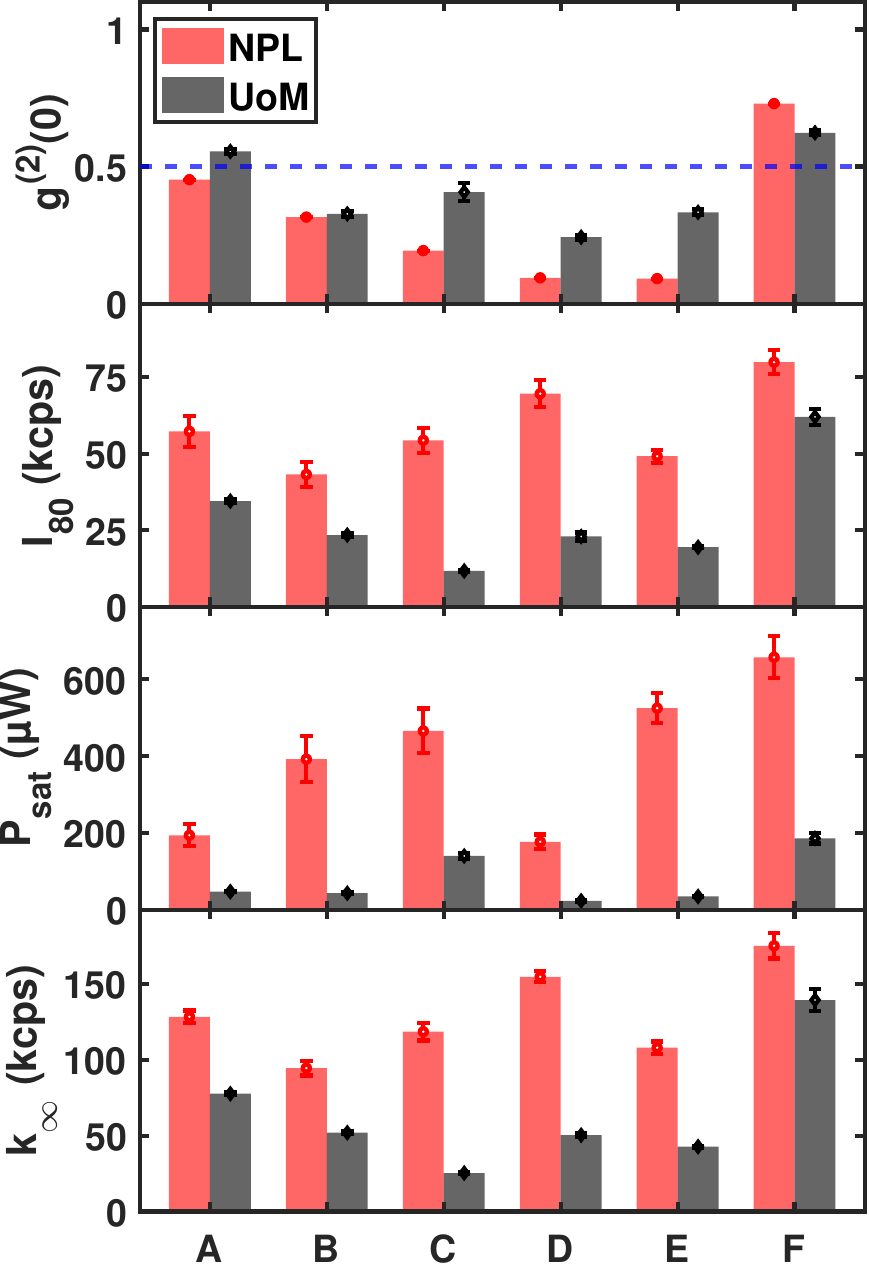}
    \caption{Parameters of candidate NV$^{-}$ centers \#A-\#F correlated between the two systems, NPL (red) and UoM (gray). The blue dashed line represents $g^{(2)}(0)=0.5$.}
    \label{figure:parameter_plots}
\end{figure}

The best 6 potentially single emitters were selected by NPL; these had an RMSE~$<0.15$ and a sufficient SNR to suggest it is a likely SPE with $g^{(2)}(0)<0.5$. These candidates were located at UoM by referencing fiducial markers on the substrate (see~\SI). They were examined under the same excitation conditions as NPL as demonstrated in FIG. \ref{figure:tcspc_mapping}(b), meaning the absorbed fluence was identical, and optoelectronic properties could be compared. In FIG. \ref{figure:parameter_plots}, the three model parameters ($g^{(2)}(0)$, $P_{sat}$, $k_{\infty}$) and $I_{80}$ are compared between NPL and UoM, where error bars represent a 1$\sigma$ uncertainty in the fitting parameter. The $g^{(2)}(0)$ values are expected to be independent from experimental variables, except for laser scatter and detector dark counts. Four NV$^{-}$ centers were found to have $g^{(2)}(0)<0.5$ (\#B, \#C, \#D, \#E) with the remaining 2 having $g^{(2)}(0)>0.5$ (\#A, \#F). $k_{\infty}$ is a product of the fundamental emission rate and system throughput, and therefore the ratio of this parameter between NPL and UoM is sensitive to these differences. The ratio of $I_{80}$ between NPL and UoM is also sensitive to variation in the system throughput, as well as the background count rate. Conversely, the target pump power required to reach this count rate depends on the light coupling into the NV$^{-}$ center, which is primarily sensitive to the excitation beam profile. This is shown by table \ref{tab:ND5_repeats}, where $P_{sat}$ changes between system calibrations.

The uncertainties in FIG. \ref{figure:parameter_plots} are predominantly attributable to the SNR, which can be improved by acquiring photon streams for a longer duration. However, excitation conditions and polarization may also play a role. Excitation conditions, such as the wavelength, can affect recombination lifetimes\cite{Ji2018Multiple-photonDiamond} which changes the appearance of the $g^{(2)}(t)$ signal. However, this is not expected to be significant in this work as both systems use nominally identical pump wavelengths. The orientation of the NV$^{-}$ center dipoles determine the perceived emission rate, by affecting how efficiently the color center is excited by polarized light,\cite{Dolan2014CompleteBeams} causing a change in $P_{sat}$, and the emission profile of the NV$^{-}$ center. Since the candidates have high emission rates in both NPL and UoM systems, despite sample orientation not being controlled, this indicates that these particular NDs may host favorably oriented emission dipoles resulting in no overall polarization for absorption or emission where the emission dipoles lay parallel to the focal plane.\cite{Dolan2014CompleteBeams} This is a favorable characteristic for a reference single-photon emitter, as control for polarization is not required. Therefore while an SPE that emits with $g^{(2)}(0)<0.5$ can be used as an SPE reference, the closer the $g^{(2)}(0)$ value is towards zero, the less variability is expected in emission. Non-single-photon processes such as scattering likely contribute to increased variability in emission rate because, as purely optical effects, they are sensitive to ND orientation and laser pump profile.

Critically, the dissimilarities in throughput reveals a systematic difference in observed properties between laboratories and explains why $I_{80}$, $P_{sat}$ and $k_{\infty}$ in FIG. \ref{figure:parameter_plots} are expected to lie outside their uncertainty bounds. The pump power density can change with beam alignment between and over the duration of the measurement which is reflected by $P_{sat}$ in table \ref{tab:ND5_repeats}, to reach a $k_{\infty}$ of 52(1)~kcps. This indicates a 3\% standard deviation in the determined light throughput at UoM.

\begin{figure}
    \centering
    \includegraphics[width=8.5cm]{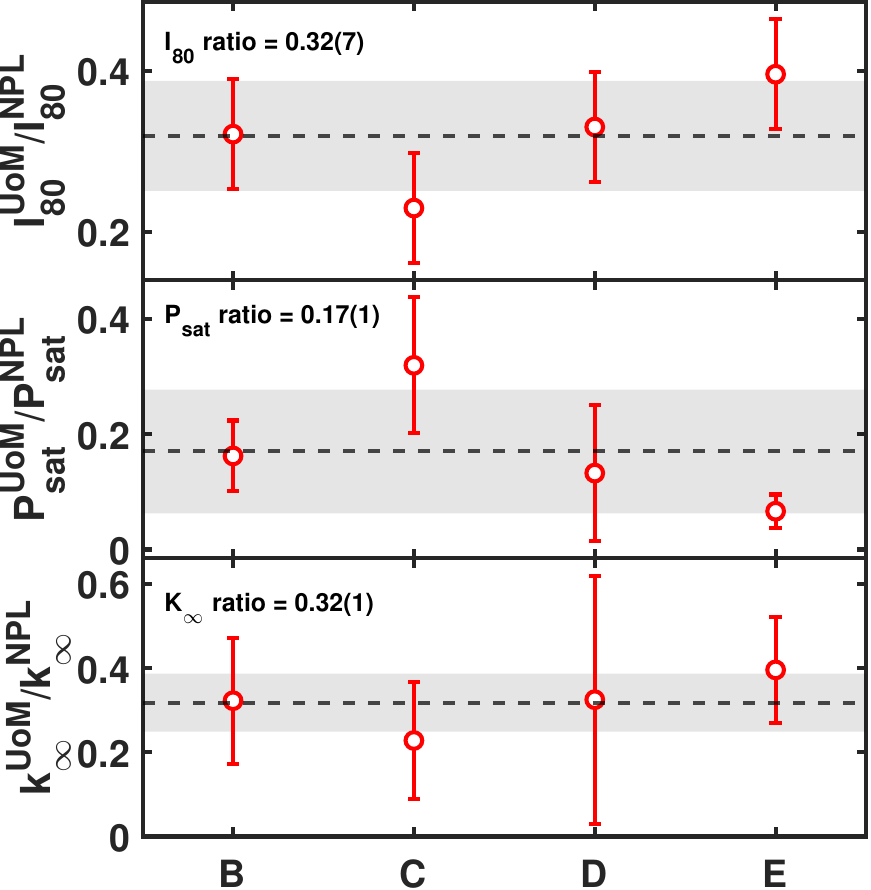}
    \caption{Ratios of parameters $I_{sat}$, $P_{sat}$ and $k_{\infty}$ between NPL and UoM at saturation for single-photon emitting NV$^{-}$ centers. The error bars represent the ±1$\sigma$ fitting error for each center, the horizontal dashed black line indicates the mean across these four centers, and shaded regions shows the experimental ±1$\sigma$ standard deviation across the four centers.}
    \label{figure:ratios}
\end{figure}

The UoM:NPL ratio of $I_{80}$ in FIG. \ref{figure:ratios} shows that the collection efficiency at UoM is lower than NPL by a ratio of 0.32(7), when only considering single emitters \#B, \#C, \#D and \#E. Assuming Lambertian emission, the ratio (UoM:NPL) of light gathering ability solely due to the objective numerical apertures is 0.94, a minor contribution to the overall ratio of $I_{80}$. Although a multimode fiber was used at UoM, which likely collects more background photons than a single-mode fiber at NPL, the photon detection rate at $I_{80}$ was still lower at UoM. Given this approximately 3$\times$ reduction in photon detection rate, a 9$\times$ increase in acquisition time is required to maintain the same coincidence count. This is reflected in the 3600~s acquisition time for $g^{(2)}(t)$ at UoM, where a ratio of 6$\times$ was used that was sufficient for consistent results as shown in table \ref{tab:ND5_repeats}. The 0.2(1) ratio for $P_{sat}$ shows that the excitation efficiency is greater at UoM, with an intrasite standard deviation of 35.8\%. Systematic differences between NPL and UoM are shown by the standard deviation of 30.35\% in $I_{80}$ and 60.17\% in $P_{sat}$, suggesting that calibration is required for quantitatively comparing inter-site single-photon measurements.

\begin{table}
\centering
    \caption{$g^{(2)}(0)$, $I_{80}$, $I_{sat}$, $P_{sat}$ and $k_{\infty}$ values upon repeated measurement of ND \#B at UoM (UoM$_{1-3}$) with reference values measured at NPL (NPL$_{1}$). The errors represent a fitting $\pm$1~$\sigma$ standard deviation.}
    \begin{tabular}{cccccc}
    \hline\hline
        Repeat & $g^{(2)}(0)$ & $I_{80}$ (kcps) & $I_{sat}$ (kcps)& $P_{sat}$ (${\mu}$W)& $k_{\infty}$ (kcps)\\
        \hline
        NPL$_{1}$ & 0.32(3) & 43(4) & 47(2) & 390(60) & 94(5)\\
        UoM$_{1}$ & 0.33(1) & 23.4(6) & 26.0(5) & 43(2) & 52(1)\\
        UoM$_{2}$ & 0.20(1)& 22.5(6) & 25.1(5) & 60(2) & 50(1)\\
        UoM$_{3}$ & 0.31(1) & 24.4(8) & 26.6(8) & 89(5) & 53(2)\\
        \hline\hline
    \end{tabular}
    \label{tab:ND5_repeats}
\end{table}

Initial measurements on the 6 candidates showed that emitter \#B, shown in FIG. \ref{figure:tcspc_mapping}(a), was the least sensitive to system alignment, possibly due to favorable orientation of the NV$^-$ center. This candidate gave an $I_{80}$ count rate of 23(1) kcps, as calculated from table \ref{tab:ND5_repeats}, showing a 3\% standard deviation upon repeated measurements, which serves as an exemplary SPE for this work. FIG. \ref{figure:tcspc_mapping}(c) shows that for ND \#B the $g^{(2)}(t)$ signal measured at both laboratories are in agreement, with $g_{UoM}^{(2)}(0)=0.28(7)$ compared to $g_{NPL}^{(2)}(0)=0.32(3)$. We note that the $g^{(2)}(0)$ of 0.20(1) in table \ref{tab:ND5_repeats} appears to be an outlier, however, it still confirms that ND \#B is an SPE. The sample was repeatedly studied over 3 months (see SI), ND \#B was irradiated several times with power densities up to 2.53~MW/cm$^{2}$ for up to 3600~s at a time. No optically induced damage was observed. This is compatible with previous studies, which have reported stability that is expected to last for centuries,\cite{Laube2023} demonstrating the robustness of the NV$^{-}$ centers hosted in a ND system.

Another color center, \#C, did not exhibit any metastable state decay times in its $g^{(2)}(t)$ (see~\SI) which is ideal for generation of on-demand single photons,\cite{Eisaman2011} as metastable states\cite{Li2015LifetimeNanodiamonds} would otherwise reduce the rate of radiative recombination.  This could be due to the local environment of the NV$^{-}$ center, where there is an absence of other nitrogen defects in the near vicinity that allow a photoinduced electron transfer to enable charge state switching.\cite{Gaebel2006} This would result in a lack of intersystem crossings, and therefore no metastable state decay as the system is prevented from doing so. We note that such NV$^{-}$ centers are rare, representing less than 0.1\% of those studied in the present set.

Although the robust and stable nature of NV$^{-}$ centers as SPE references has been demonstrated, it is not without caveats. Firstly, it was inefficient to obtain enough candidate reference emitters as only 4 NDs (66.7\%) were validated to contain NV$^{-}$ centers at UoM and only 1 (16.7\%) showed repeatable results. The high-throughput data (see~\SI)~follows a similar trend where 44\% of NDs contain single defects,\cite{Wood2022LongNanodiamonds} yet few are usable as quantum references due to poor signal-to-noise ratio, which is why a subset was selected for remeasurement. An alternative would be to use defect implantation\cite{Adshead2023AEngineering,Rabeau2006Implantation15N,Orwa2011EngineeringAnnealing,Chu2014CoherentCenters,Meijer2005GenerationImplantation} for greater specificity over defect concentration, although these techniques are expensive in time and cost. Another approach is to increase the efficiency of the procedure through QR codes etched on the substrate.\cite{Sutula2023} This would allow machine-assisted location of NDs by automatically calculating the rotation of the sample and routing to the unique code hash closest to the candidate, improving the reliability of location for repeated measurement. Another limitation is that the NV$^{-}$ centers are most efficiently excited by 510~nm to 575~nm radiation,\cite{Aslam2013Photo-inducedDetection,Beha2012OptimumDiamond} confining other measurements to materials that are excited in this wavelength range. It is important to note that the NV centers are sensitive to changes in magnetic field,\cite{Gruber1997} and so this approach is most applicable in the absence of external fields.

\section{\label{sec:Conclusion}Conclusion}
Since single-photon characterization systems generally differ in both collection and excitation efficiency, robust SPE references are needed for calibration. We have demonstrated that, by measuring the count-rate saturation curves of specific NV$^{-}$ centers, it is possible to standardize the excitation conditions across different experimental setups, facilitating reliable and repeatable measurements of the single-photon emission rate. The $I_{80}$ count rate ratio of 0.32(7) between UoM:NPL, and the $k_{\infty}$ ratio of 0.32(1) indicates that this technique can be used to correct for a system's photon throughput. Crucially, when used as a benchmark for other materials, this allows quantitative comparison to externally measured materials using the same technique, with their own NV$^{-}$ reference. Our cross-site calibration shows that it is possible to use locally identified single NV- centers as a single-photon reference which is portable, stable and robust. This shows promise as a method to standardize comparison of a wide range of quantum materials, supporting work towards realizing deterministic absolute single-photon sources.

\begin{backmatter}
\bmsection{Funding}
N.P. acknowledges the EPSRC (UK) for a studentship. P.P. acknowledges the funding from InnovateUK Commercialising Quantum Technologies [TS/X002195/1] and the UKRI Future Leaders Fellowship [MR/T021519/1]. S.C. acknowledges funding from the Leverhulme Trust [ECF-2024-250]. NPL acknowledges the support of the UK government department for Science, Innovation and Technology through the UK national quantum technologies program.

\bmsection{Acknowledgment}
We would like to acknowledge Himanshu Shekha and Nu Quantum for provision of registration featured substrates. We thank Dr Hannah Stern (UoM) for her feedback on an early draft of this work, and Dr Christopher Chunnilall (NPL) for his useful revisions and contributions to several drafts of this manuscript.

\bmsection{Author Contributions}\\
\textbf{Nikesh Patel}: Methodology (lead), Software (supporting), Validation (equal), Formal Analysis (lead), Investigation (equal), Data curation (lead), Writing - Original Draft (lead), Writing  Review \& Editing (equal), Visualization (lead). \\
\textbf{Benyam Dejen}: Methodology (lead), Software (supporting), Validation (equal), Formal Analysis (supporting), Investigation (equal), Data curation (supporting), Writing  Review \& Editing (equal). \\
\textbf{Dr Stephen Church}: Methodology (supporting), Software (lead), Writing - Review \& Editing (equal). \\
\textbf{Dr Philip Dolan}: Conceptualization (equal), Methodology (lead), Software (lead), Validation (equal), Formal analysis (supporting), Resources (lead), Writing - Review \& Editing (equal), Supervision (equal), Project administration (equal), Funding acquisition (equal). \\
\textbf{Dr Patrick Parkinson}: Conceptualization (equal), Methodology (lead), Software (lead), Validation (equal), Formal analysis (supporting), Resources (lead), Writing - Review \& Editing (equal), Visualization (supporting), Supervision (equal), Project administration (equal), Funding acquisition (equal).%

\bmsection{Disclosures}
The authors declare no conflicts of interest.

\bmsection{Data Availability Statement}
The data that support the findings of this study are openly available in Figshare at http://doi.org/10.48420/27290712. The code associated with the analysis is available at https://github.com/OMS-lab/Nanodiamond-Nitrogen-Vacancy-Reference-Emitters.

\bmsection{Supplemental document}
See Supplement 1 for supporting content. This includes fiducially marked substrates, experimental systems, NV$^{-}$ center identification, high-throughput statistics, normalized saturation curve for ND \#B, pulsed g$^{(2)}$(t) measurement on ND \#B, A-F candidate $g^{(2)}(t)$ signals, power dependent photoluminescence spectra for ND \#B, ND \#B Count rate variation with time and a comparison of measured parameters in this study with those in the literature for NV centers and other single photon sources.

\end{backmatter}


\bibliography{references}

\end{document}


\maketitle

\section{Fiducially Marked Substrates}
\begin{figure}
    \centering
       \includegraphics[width=\textwidth]{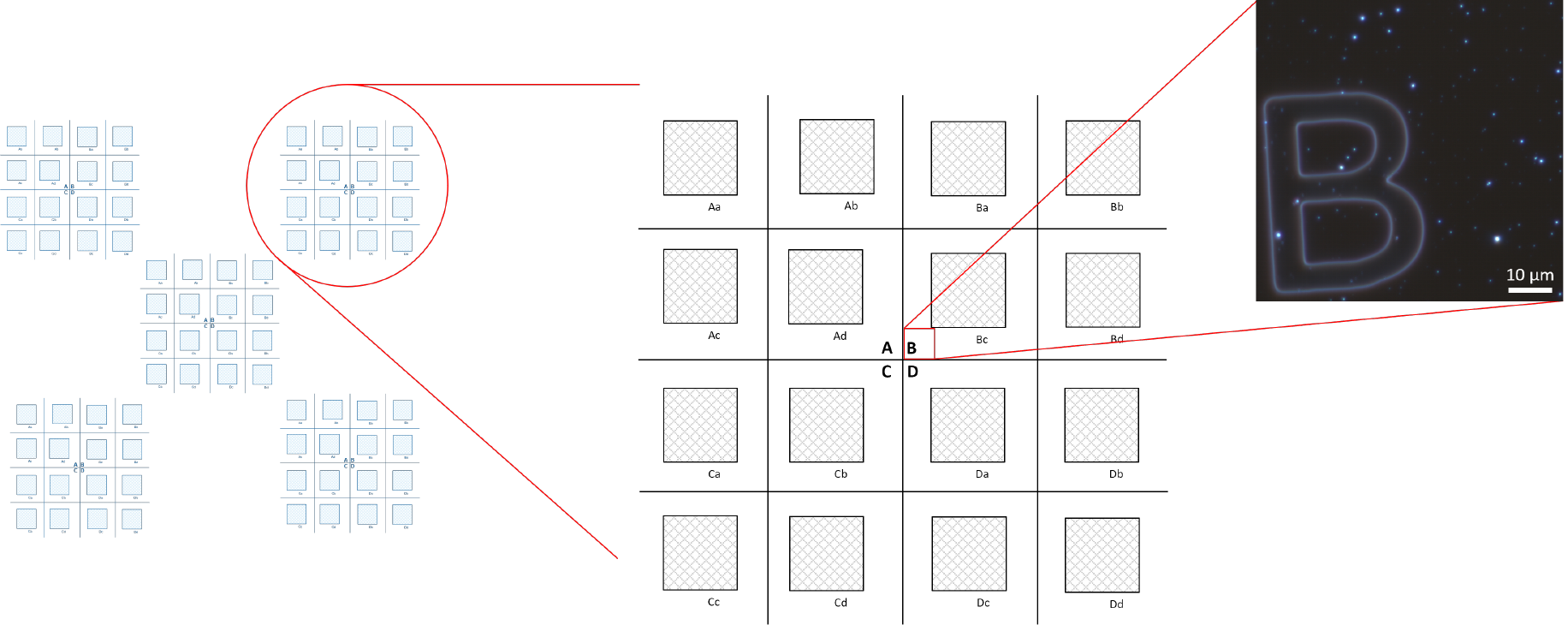}
    \caption{Diagram of the marker pattern etched on the silicon substrate, spanning a $2\times2$~mm area. A dark-field microscope image of an etched uppercase B is included to the right, with regions of high emission intensity representing NDs.}
    \label{figure:marked_substrate}
\end{figure}

Identifying specific single objects with sizes of the order of microns to hundreds of nanometers is extremely difficult without reference markers that are invariant under rotation, translation and reflection. This can be achieved using a QR code system for machine readable referencing or human-readable markers as shown in this experiment. The inset in FIG. \ref{figure:marked_substrate} shows an optical microscope image that was used for calculating the density of nanodiamonds per unit area. This was done by thresholding the image to reject low intensity background, running an object detection algorithm (MATLABs \textit{regionprops}), filtering results down to point source emitters and summing the detected bright regions in the sampled area. This data was then used to calculate an areal density of NDs of 0.03~$\mu$m$^{-2}$.

A silicon substrate was etched with markers that are invariant under rotation and translation with patterns on the order of 20-500 $\mu$m. The marked references included combinations of the letters A, B, C and D in upper- and lower-case, along with a grid pattern shown in FIG. \ref{figure:marked_substrate}. Following the etch, NDs were spincast from solution on to this substrate.

\section{Experimental Systems}
\subsection{NPL}
\begin{figure}
    \centering
       \includegraphics[width=\textwidth]{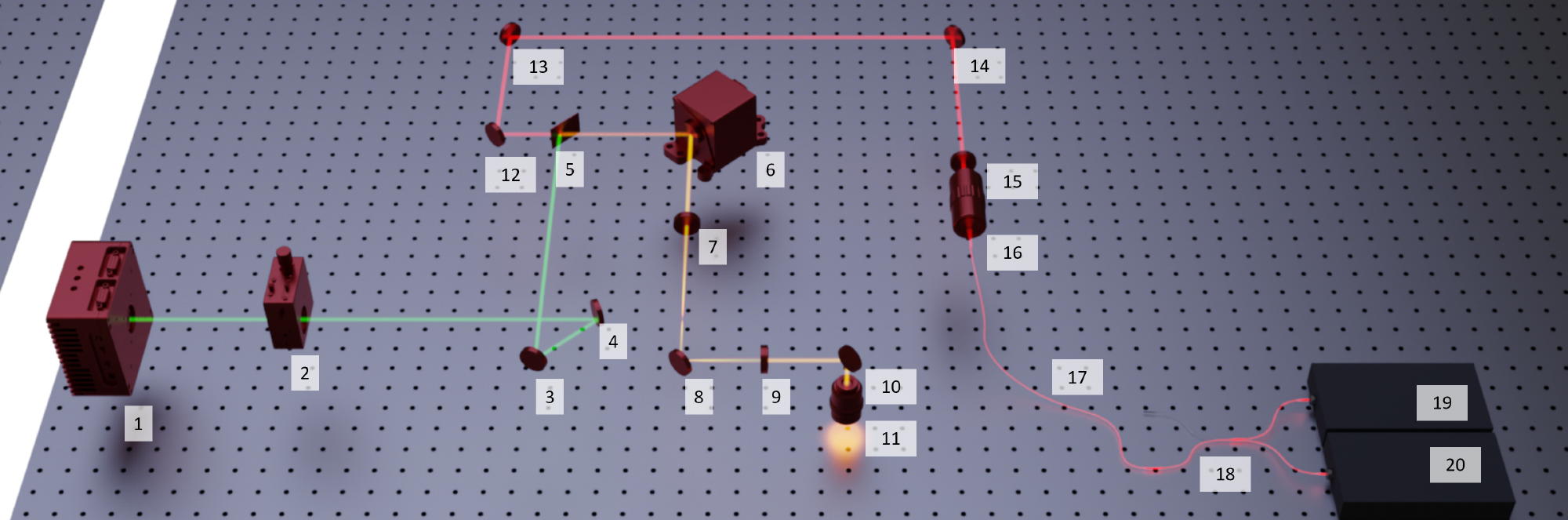}
    \caption{Diagram of the layout used for high-throughput identification at NPL.}
    \label{figure:npl_setup}
\end{figure}

The NPL set-up is illustrated in Fig. \ref{figure:npl_setup}, with the components listed in Table \ref{tab:npl_expt}.

\begin{table}
    \centering
    \resizebox{\columnwidth}{!}{%
    \begin{tabular}{cc}
    \hline
        Label & Equipment\\
    \hline
    \hline
        1 & Thorlabs DJ532-40 532 nm, 40 mW, E Pin Code, DPSS Laser\\
        2 & Thorlabs NEL02A/M High-Power Noise Eater\\
        3 & Thorlabs BB1-E02 Ø1" Broadband Dielectric Mirror, 400 - 750 nm\\
        4 & Thorlabs BB1-E02 Ø1" Broadband Dielectric Mirror, 400 - 750 nm\\
        5 & Thorlabs DMLP550R 25 mm x 36 mm Longpass Dichroic Mirror, 550 nm Cut On\\
        6 & Newport FSM-300-01 Fast Steering Mirror System\\
        7 & Thorlabs AC254-125-A-ML f=125 mm, Ø1" Achromatic Doublet, SM1-threaded Mount, ARC: 400-700 nm\\
        8 & Thorlabs BB1-E02 Ø1" Broadband Dielectric Mirror, 400 - 750 nm\\
        9 & Thorlabs AC254-125-A-ML f=125 mm, Ø1" Achromatic Doublet, SM1-threaded Mount, ARC: 400-700 nm\\
        10 & Thorlabs BB1-E02 Ø1" Broadband Dielectric Mirror, 400 - 750 nm\\
        11 & Olympus UPLFLN 60X Objective, Edmund Optics\\
        12 & Thorlabs BB1-E02 Ø1" Broadband Dielectric Mirror, 400 - 750 nm\\
        13 & Thorlabs BB1-E02 Ø1" Broadband Dielectric Mirror, 400 - 750 nm\\
        14 & Thorlabs BB1-E02 Ø1" Broadband Dielectric Mirror, 400 - 750 nm\\
        15 & Thorlabs FELH0550 Ø25.0 mm Premium Longpass Filter, Cut-On Wavelength: 550 nm\\
        16 & Thorlabs MY10X-823 10X Mitutoyo Plan Apochromat Objective, 480 - 1800 nm, 0.26 NA, 30.5 mm WD\\
        17 & Thorlabs P3-780A-FC-2 Single Mode Patch Cable, 780 - 970 nm, FC/APC, Ø3 mm Jacket, 2 m Long\\
        18 & Thorlabs TM50R5F2B 2$\times$2 Multimode Fiber Optic Coupler, Low OH, Ø50 µm Core, 0.22 NA, 50:50 Split, FC/PC\\
        19 & Excelitas SPCM-AQR-14-FC, Single-Photon Counting Module, Silicon Avalanche Photodiode\\
        20 & Excelitas SPCM-AQR-14-FC, Single-Photon Counting Module, Silicon Avalanche Photodiode\\
    \hline
    \end{tabular}
    }
    \caption{Experimental equipment used at NPL. The sample sits below item \#11.}
    \label{tab:npl_expt}
\end{table}

\subsection{UoM}
\begin{figure}
    \centering
       \includegraphics[width=\textwidth]{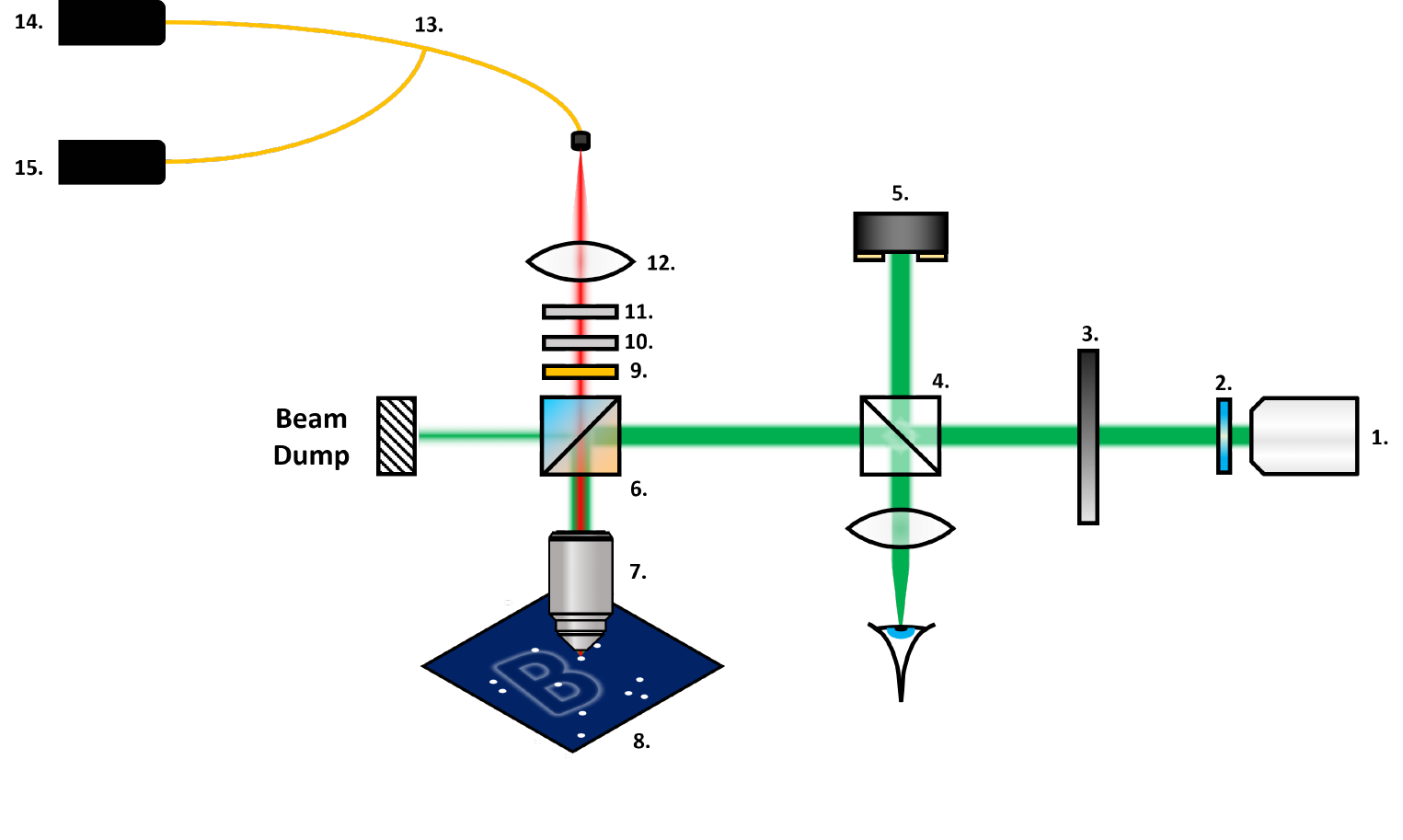}
    \caption{Diagram of the layout used for high-throughput identification at UoM. Laser emission at 532~nm (green) and NV$^{-}$ emission at 637~nm (red).}
    \label{figure:uom_setup}
\end{figure}

The NPL set-up is illustrated in FIG. \ref{figure:uom_setup}, with the components listed in Table \ref{tab:uom_expt}.

The NDs were searched for manually using optical images from a confocal microscope system using the fiducial markers as a guideline. A preliminary count rate map was run to inspect the local region to find the emitting candidates in the HBT system. The NV$^{-}$ centers were then matched to a map provided by NPL to identify the candidates. Power saturation and TTTR photon streams were then acquired by dwelling on the candidate for 3600\,seconds.

\begin{table}
    \centering
    \resizebox{\columnwidth}{!}{%
    \begin{tabular}{cc}
    \hline
        Label & Equipment\\
    \hline
    \hline
        1 & 532 nm Ventus DPSS continuous wave laser\\
        2 & Thorlabs FLH05532-4 532nm laser line filter\\
        3 & Thorlabs NDC-100C-2M Graded ND filter\\
        4 & Thorlabs BS013 50:50 beamsplitter\\
        5 & Thorlabs PM120VA power meter\\
        6 & Thorlabs DMLP550R Dichroic filter\\
        7 & Leica 566073 100$\times$, 0.85 NA objective\\
        8 & Sample mounted on Physik Instrumente V-738 High Precision XY stage\\
        9 & Thorlabs FELH0550 550 nm long pass filter\\
        10 & Horiba XE532 532 nm edge Raman filter (OD 10$^5$)\\
        11 & 2" 532 nm Raman notch filter (OD 10$^7$)\\
        12 & Thorlabs AC254-200-AB best form focusing lens (f = 40 mm)\\
        13 & Thorlabs TM50RF1B 50:50 beamsplitter multimode fiber\\
        14 & ID Quantique ID100 SPAD\\
        15 & ID Quantique ID100 SPAD\\
    \hline
    \end{tabular}
    }
    \caption{Experimental equipment used at UoM.}
    \label{tab:uom_expt}
\end{table}

The main differences between the two systems are the NA of the objectives, collection lens for focusing light on to the collection point fiber, and the method of power attenuation (liquid crystal noise eater vs neutral density filters). The TCSPC modules are not shown in the diagrams, but are connected directly to the SPADs via BNC cables.

\section{NV\texorpdfstring{$^{-}$}~ Center Identification}
\begin{figure}
    \centering
       \includegraphics[width=\textwidth]{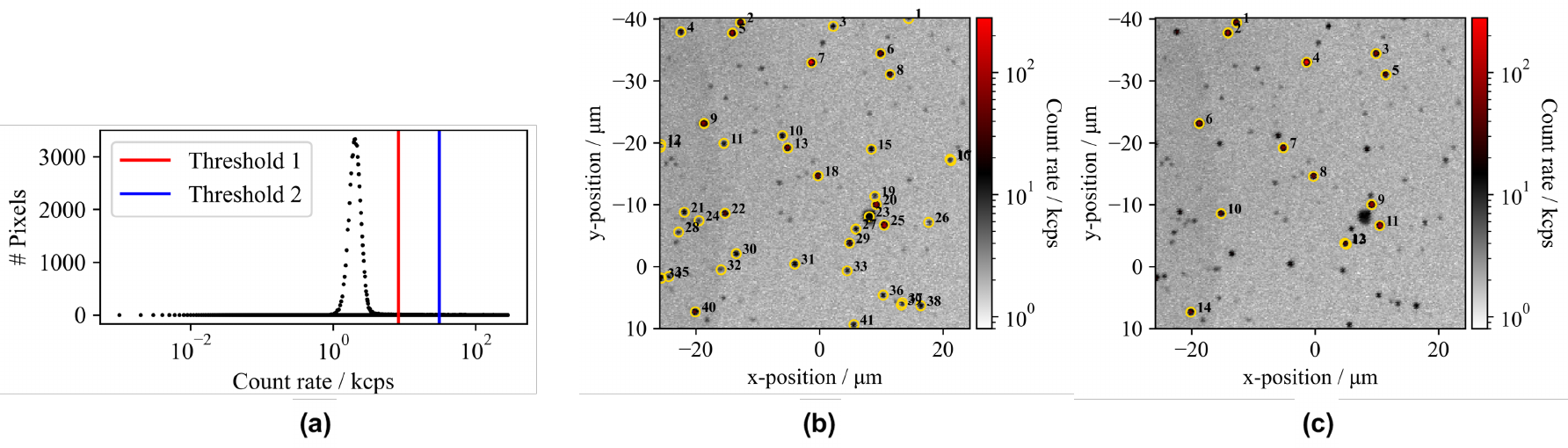}
    \caption{a) Histogram of count rates (log scale) from a TCSPC mapping, indicating the mode value and user-defined thresholds 1 and 2. b) Object identification after applying threshold 1, $T_{1}$. c) Objects remaining after applying threshold 2}
    \label{figure:npl_algo}
\end{figure}

The object finding algorithm used by NPL first generates a histogram of the  count rates, $I$, from each pixel in the 2D raster scan search ($256\times256$ pixels spaced at 200~nm covering an area approximately $50\times50~\mu$m) , as shown by FIG. \ref{figure:npl_algo}(a). The mode value, $I_{mode}$ of this is taken as representation of the background count rate of the substrate, and a threshold is set above this, by $T_{1}=x_{1}I_{mode}$, where the user-specified factor $x_{1}>1$. $x_{1}$ is defined through trial and error to identify regions of high emission intensity. The mapping is masked by setting pixels with $I<I_{mode}=0$, and convolved with a $10\times10$~pixel 2D Gaussian to smooth out the objects. The coordinates of the resulting bright spots are retrieved using the publicly available Python package \textit{scipy.ndimage.find\_objects}, and are marked as shown in FIG. \ref{figure:npl_algo}(b). Since there are numerous objects, the identified objects are filtered by setting a second threshold $T_{2}=x_{2}I_{mode}$ with the user-defined factor $x_{2}>x_{1}$ to obtain a subset of the highest intensity emitting objects as shown in FIG. \ref{figure:npl_algo}(c). Objects with greater emission intensity are likely to have improved signal-to-noise ratio, increasing the chances of finding a suitable candidate for remeasurement.

\section{High-throughput Statistics}
\begin{figure}
    \centering
       \includegraphics[width=0.75\columnwidth]{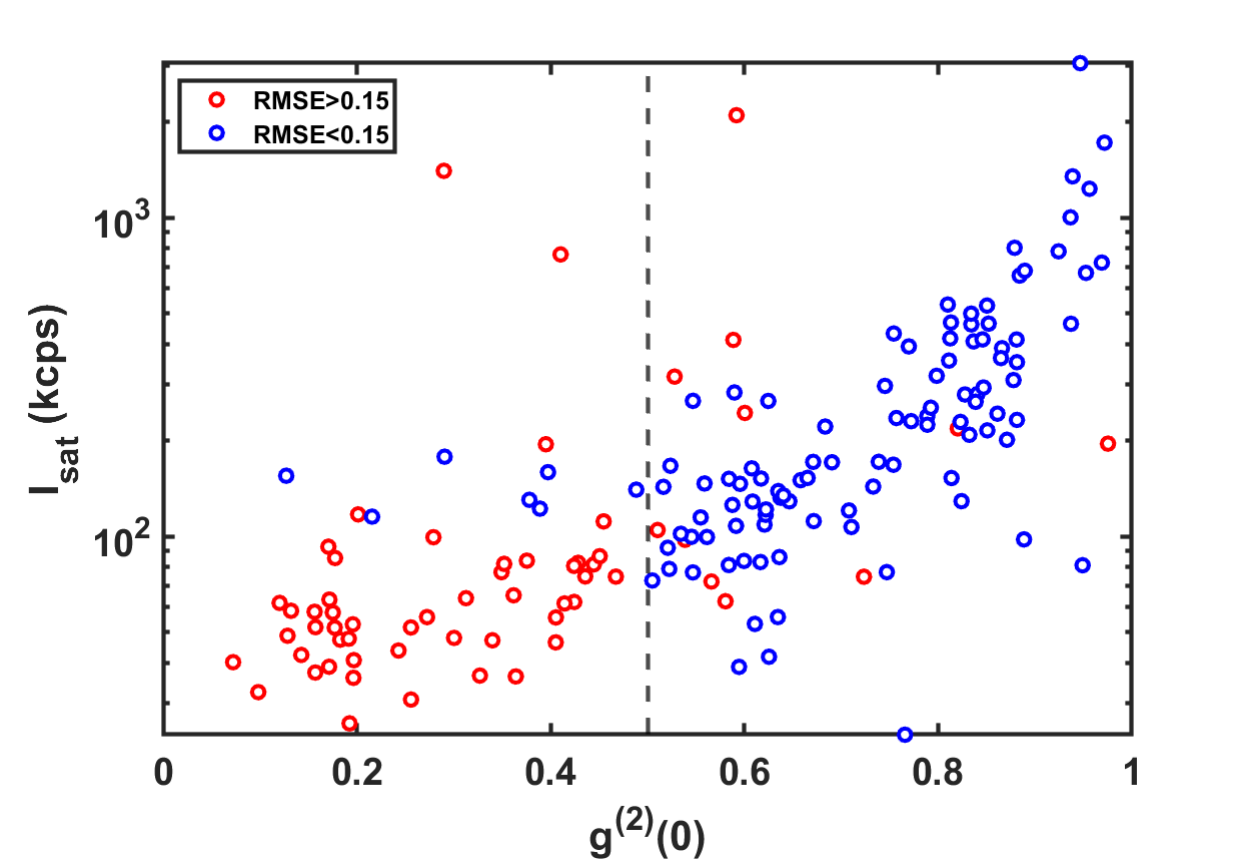}
    \caption{High-throughput stats from the identification stage of finding single NV$^{-}$ centers at NPL. Data points with $g^{(2)}(0)=0$ are filtered out and $I{sat}<5$~kcps are excluded to remove data originating from fits with poor SNR.}
    \label{figure:npl_stats}
\end{figure}

\begin{figure}
    \centering
       \includegraphics[width=0.75\columnwidth]{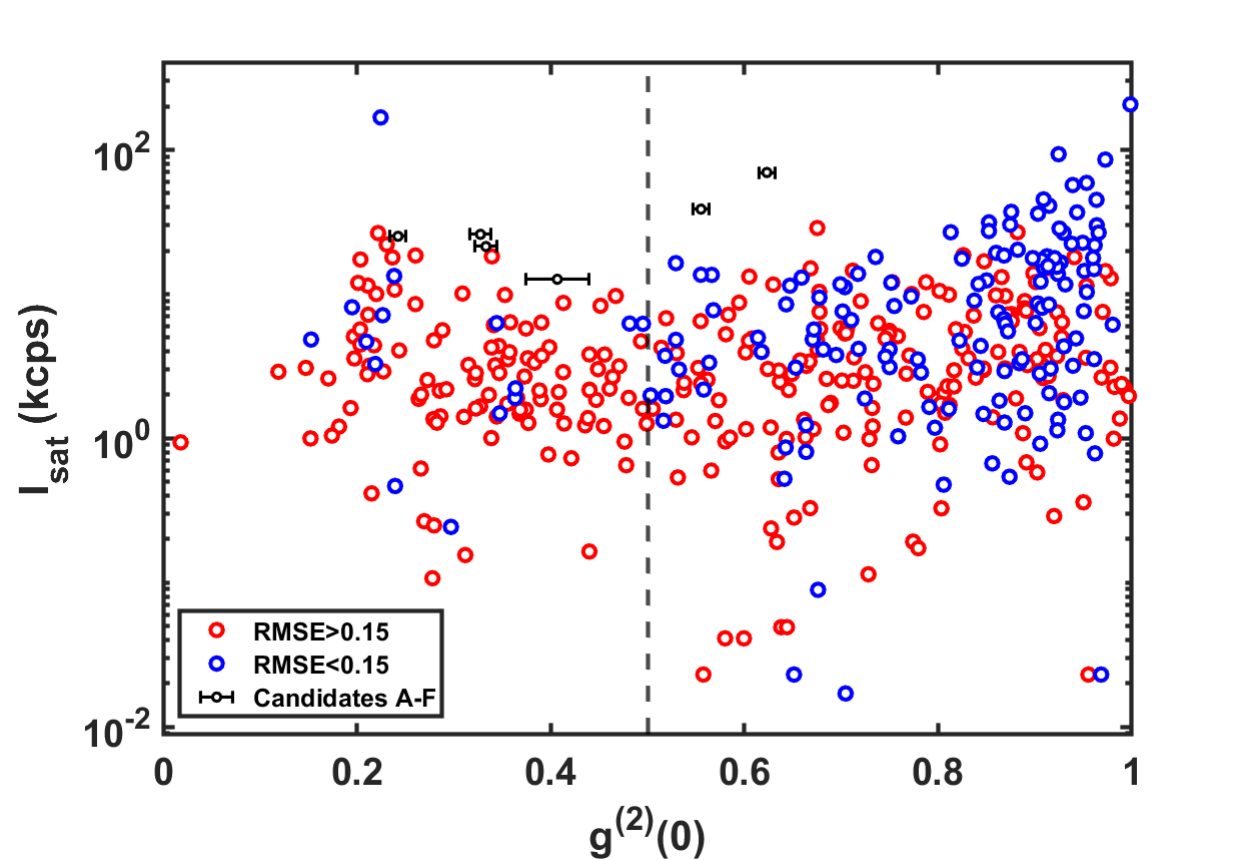}
    \caption{High-throughput data from UoM, the candidates were measured after swapping the collection lens from $f=20$\,mm to an $f=40$\,mm lens for improved light collection, which is reflected by the higher  count rate relative to the rest of the data.}
    \label{figure:uom_stats}
\end{figure}

High-throughput data shown in FIGs. \ref{figure:npl_stats} and \ref{figure:uom_stats} highlight that although there are several NV$^{-}$ centers with $g^{(2)}(0)<0.5$, few are selected as candidates as a manual quality control step is applied after algorithmic identification of possible candidates. The algorithm was susceptible to identifying remeasurement candidates with poor $g^{(2)}(t)$ fits due to low signal-to-noise ratio (SNR) around $g^{(2)}(0)$. A few NV$^{-}$ centers were defined as having $g^{(2)}(0)=0$, a false reading due to poor fitting and associated with an RMSE$>0.15$, therefore was filtered out from algorithmic candidate selection.

\section{Normalized Saturation Curve for ND \#B}
\begin{figure}
    \centering
       \includegraphics[width=0.5\columnwidth]{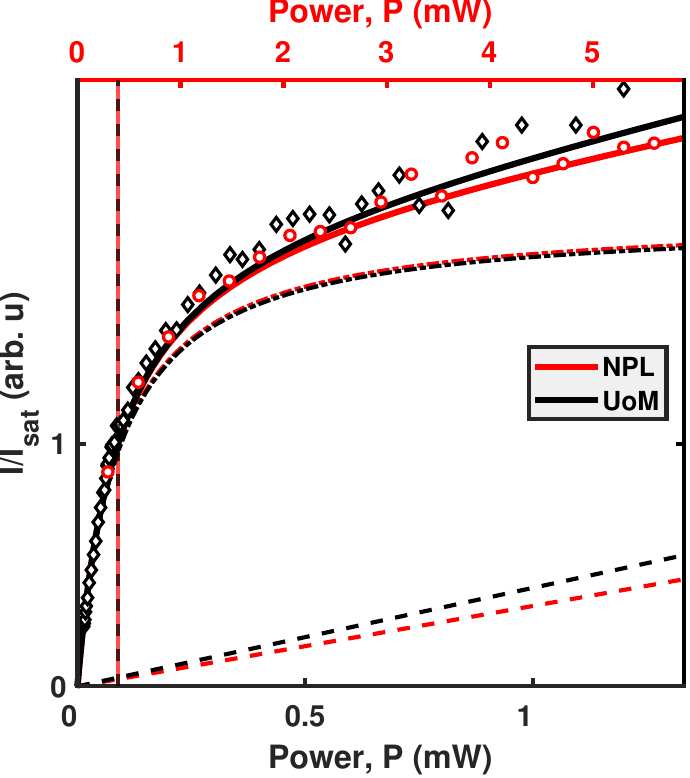}
    \caption{NV saturation curves (dashed-dot) normalized to respective I$_{sat}$ values closely overlap, supporting that the NV center is independently probed under the same excitation conditions.}
    \label{figure:NDB_normalised}
\end{figure}

In figure \ref{figure:NDB_normalised}, the saturation curve for ND \#B is normalized to I$_{sat}$ (i.e. y-values divided by respective I$_{sat}$). This shows the contributions from the NV emission (dashed-dot lines) overlap well, indicating that we are probing at the same point on the saturation curve as intended.

\section{Pulsed Measurement on ND \#B}
\begin{figure}
    \centering
       \includegraphics[width=0.7\columnwidth]{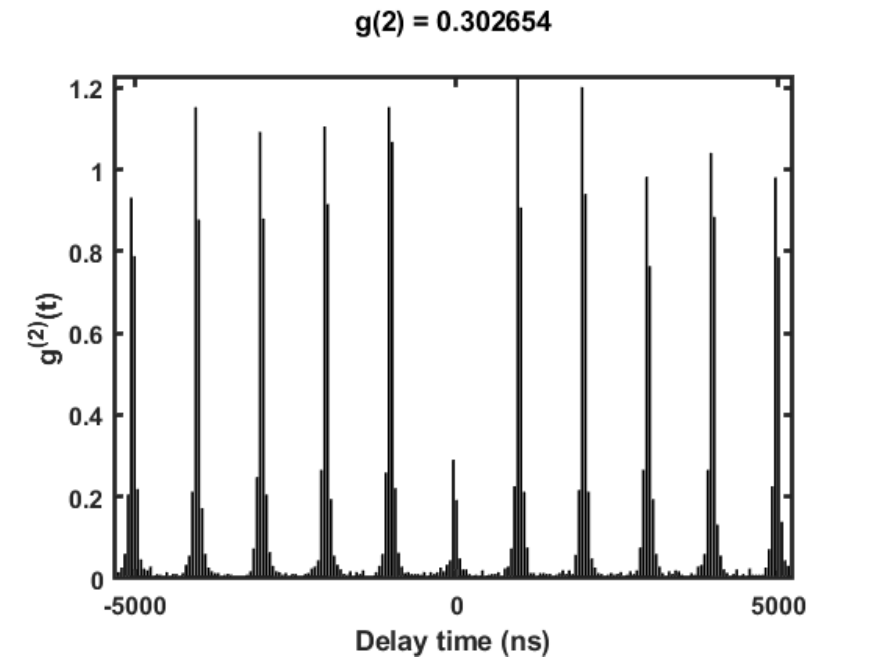}
    \caption{Pulsed g$^{(2)}$(t) measurement for ND \#B.}
    \label{figure:NDB_pulsed}
\end{figure}

To confirm the g$^{(2)}$(0) value for ND \#B, we measured this in the pulsed regime at room temperature using a 515~nm source pulsed at 1~MHz. The coincidence counts were binned into a histogram and normalized by the mean maximum frequency in a 1~$\mu$s window for pulses t~$>{\lvert}1{\rvert}~\mu$s. This gave g$^{(2)}$(0)~=~0.303 as shown in figure \ref{figure:NDB_pulsed}.

\section{A-F Candidate \texorpdfstring{$g^{(2)}(t)$}~ Signals}
Overlayed $g^{(2)}(t)$ signals between UoM and NPL, showing that the appearance of $g^{(2)}(t)$ curves are identical when measured at the same point on the saturation curve (\#A, \#B, \#C and \#F). Samples \#D and \#E were pumped above 0.8$P_{sat}$ at UoM, as their $P_{sat}$ was below the minimum pump threshold to generate sufficient SNR (as specified in the methods section)  – exhibiting an increase in metastable state transitions, as observed by the increased shoulders of $g^{(2)}(t)$ associated with bunching processes, either side (±40~ns delay time) of the $g^{(2)}(0)$ dip compared to the signal obtained at NPL. As these NV$^{-}$ centers are pumped at different points on the saturation curve, the absorbed photon flux is dissimilar and results in a discrepancy of the $g^{(2)}(t)$ signal.

A study on the power dependence of $g^{(2)}(t)$ in NV centers hosted in nanodiamonds shows that an increased pump power results in an increased height of these bunching shoulders.[a]

\begin{figure}
    \centering
       \includegraphics[width=\textwidth]{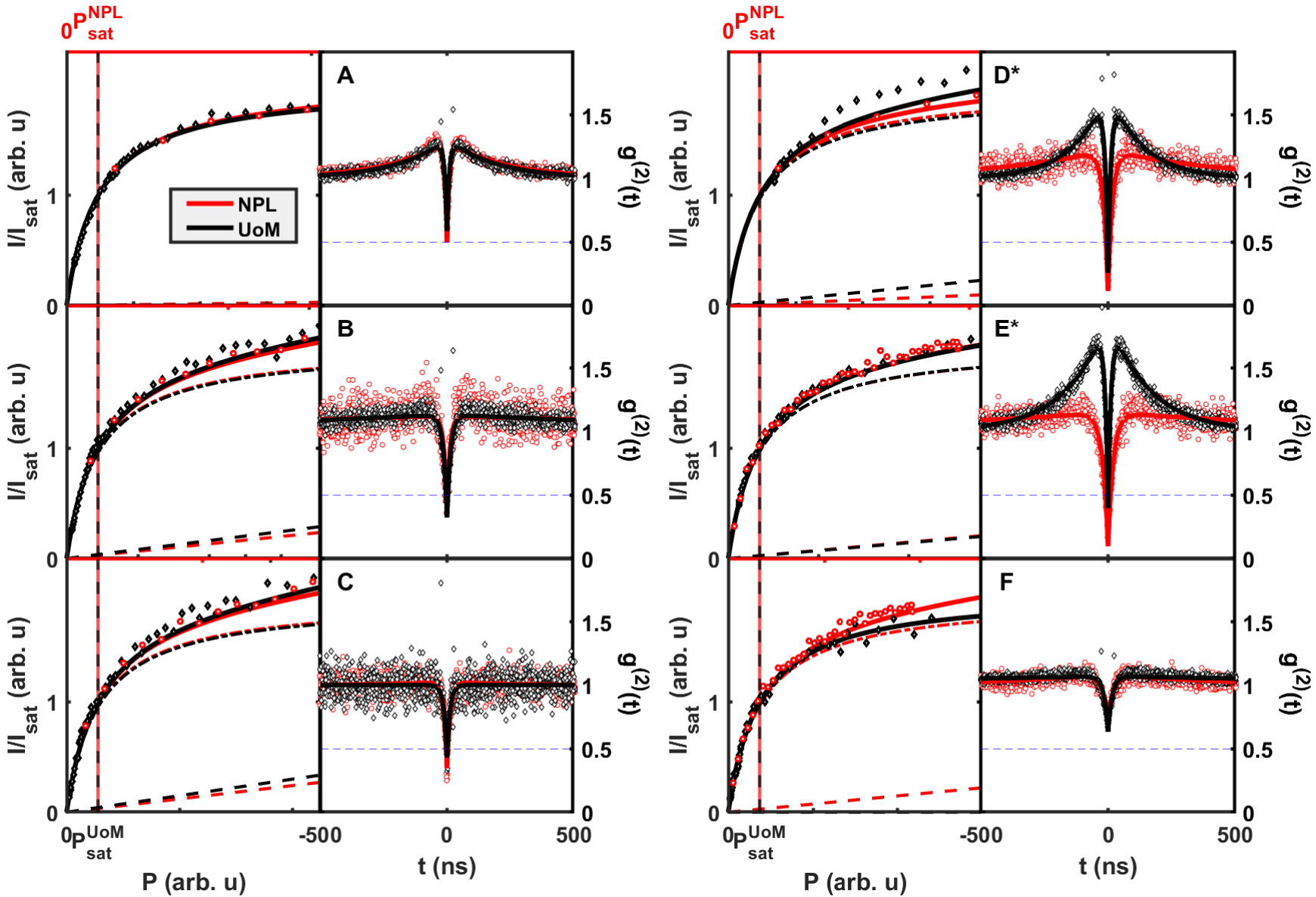}
    \caption{Normalized saturation curves and $g^{(2)}(t)$ signals for candidates A-F, UoM (black) and NPL (red). In the power dependency plots, vertical lines correspond to pump power saturation, with the full saturation curve (solid), components from the NV$^{-}$ emission (dashed-dot) and linear laser offset (dashed). The x-axis is also scaled from 0 to 8$P_{sat}$, and the curves are normalized to $I_{sat}$. (*) \#D and \#E were pumped at a minimum power threshold that exceeded 0.8$P_{sat}^{UoM}$ to achieve sufficient SNR for acquiring a g$^{(2)}$(t) signal.}
    \label{figure:AF_signals}
\end{figure}

[a] M. Berthel et al., Phys. Rev. B 91, 035308 (2015).

\section{PL Spectra for ND \#B}
\begin{figure}
    \centering
       \includegraphics[width=0.9\columnwidth]{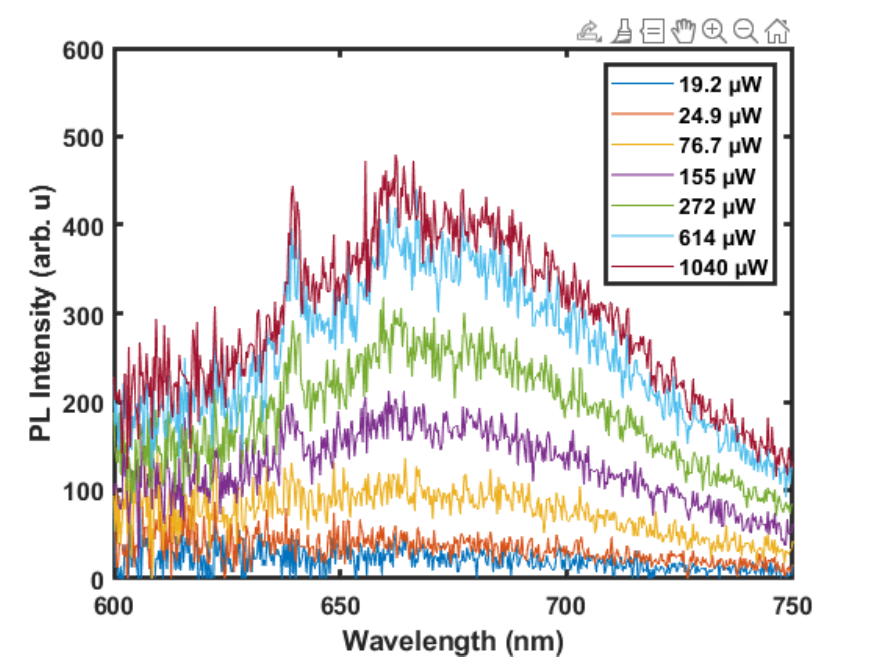}
    \caption{Interferometric TCSPC (iTCSPC) emission spectra for ND \#B accumulated over an 8 hour duration. The shoulder around 637~nm indicates the presence of an NV$^{-}$ center along with phonon sidebands at longer wavelengths.}
    \label{figure:NDB_pdep_spectra}
\end{figure}

Power-dependent photoluminescence (PL) spectra were obtained at room temperature from a Horiba ihr550 spectrometer as shown in figure \ref{figure:NDB_pdep_spectra}, showing a zero phonon line at 640~nm for ND~\#B.

\section{ND \#B Count Rate Variation}
\begin{figure}
    \centering
       \includegraphics[width=0.9\columnwidth]{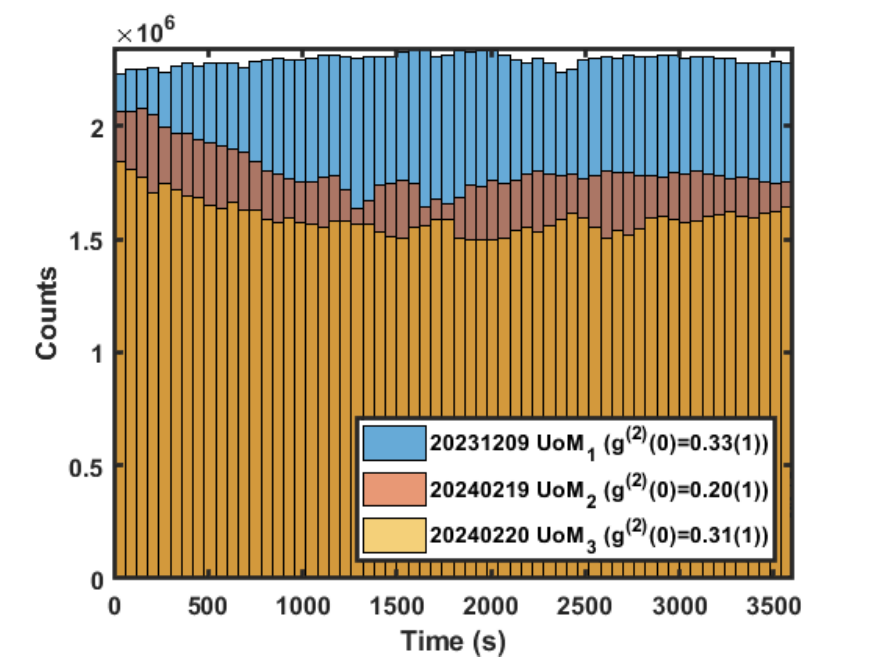}
    \caption{Count rate variation of ND \#B binned every 60~s during the 3600~s TTTR data acquisition for determining g$^{(2)}$(0).}
    \label{figure:NDB_histogram}
\end{figure}

To assess the stability of the TTTR data acquisition, a histogram of the detected photons shown in figure \ref{figure:NDB_histogram} which shows a 5-10\% variation in count rate during the course of the measurement. Sample stage drift is the likely cause of decreases in observed photon counts by around 10\% for UoM$_2$ and UoM$_3$.

The dates at which these measurements took place were as follows:
\begin{enumerate}
    \item UOM$_1$ on 09/12/2023
    \item UOM$_2$ on 19/02/2024
    \item UOM$_3$ on 20/02/2024
\end{enumerate}

This represents a 3 month period under which \#B did not photodegrade upon successive measurements.

\section{Comparison to literature values and other single photon emitters}

A selection of values for $g^{(2)}(0)$, $P_{sat}$ and $I_{sat}$ are presented in Table~\ref{tab:literature}. In particular, there is large variation in $P_{sat}$ and $I_{sat}$ between all studies, due to differences in the excitation and collection efficiency. This highlights the needed for a calibration standard, which is the primary motivation for this study.

\begin{table}
    \centering
    \resizebox{\columnwidth}{!}{
    \begin{tabular}{ccccc}
    \hline    
    SPE & Reference & $g^{(2)}(0)$ & $P_{sat}$ ($\mu$W) & $I_{sat}$ (kcps)\\
    \hline
    \hline
        NV & This work & 0.2 - 0.4 & 50 - 500 & 5 - 120 \\
        NV & [b,c] & 0.1 - 0.23 & $\approx 1000$ & 30 - 400 \\
        Defects in SiC & [d,e,f] & 0.2 & $\approx 300$ & 4 - 2000 \\      
        C nanotubes & [g] & 0.01 - 0.3 & - & 1 - 100 \\ 
        TMD QDs & [h,i,j,k] & 0.04 - 0.38 & 50 - 300 & 3 - 12000 \\ 
        Defects in hBN & [l,m] & 0.01 - 0.02 & $\approx 10000$ & $\approx 1000$ \\ 
        Colloidal QDs & [n] & 0.04 - 0.41 & - & 2 - 5 \\ 
        Perovskite QDs & [o,p,q] & 0.02 - 0.05 & $\approx 50$ & $\approx 55$ \\ 
    \hline
    \end{tabular}
    }
    \caption{Literature review of single photon sources.}
    \label{tab:literature}
\end{table}
[b] B. Rodiek, et al., Optica 4, 71 (2017). 

[c] J Aspinall, et al., Optical Materials Express 10, 2 (2020) 

[d] M. Atatüre, et al., Nat. Rev. Mater. 3, 38 (2018).   

[e] S. Castelletto, Mater. Quantum. Technol. 1, 023001 (2021).  

[f] F. Fuchs, et al., Nat Commun 6, 7578 (2015).  

[g] X. He, et al., Nat. Mater. 17, 663 (2018).   

[h] M. Von Helversen, et al., 2D Mater. 10, 045034 (2023).   

[i] Y. Luo, et al., Nat. Nanotechnol. 13, 1137 (2018).   

[j] C. Errando-Herranz, et al., ACS Photonics 8, 1069 (2021). 

[k] J. Wang, et al., Adv. Mater. 36, 2314145 (2024).  

[l] X. Li, et al., ACS Nano 13, 6992 (2019). 

[m] T. Vogl, et al., Phys Rev 3, 013296 (2021). 

[n] D. Nelson et al., J. Mater. Chem. C,12, 5684 (2024). 

[o] B.-W. Hsu, et al., ACS Nano 15, 11358 (2021). 

[p] C. Zhu, et al., Nano Lett. 22, 3751 (2022). 

[q] S. Jun, et al., ACS Nano 18, 1396 (2023).